# Parametric study of "filament and gap" models of resistive switching in TaO$_x$-based devices.

Rongchen Li, Yang Bai, Marek Skowronski*

*Department of Materials Science and Engineering, Carnegie Mellon University, Pittsburgh, Pennsylvania, 15213, USA*

ABSTRACT

A finite element model consisting of a conducting filament with or without a gap was used to reproduce behavior of TaO$_x$-based resistive switching devices. The specific goal was to explore the range of possible filament parameters such a filament diameter, composition, gap width, and composition to reproduce the conductance and shape of *I-V* while keeping the maximum temperature within acceptable range allowing for ion motion and preventing melting. The model solving heat and charge transport produced a good agreement with experimental data for the oxygen content in the filament below TaO$_{1.3}$, the filament diameter range between 6 and 22 nm, and the gap oxygen content between TaO$_{1.7}$ and TaO$_{1.85}$. Gap width was not limited on either low or high sides by the criteria considered in this report. The obtained filament composition corresponds to oxygen deficiency an order of magnitude higher than one estimated by other modelling efforts. This was in a large part due to use of recent experimental values of conductivity as a function of composition and temperature. Our modelling results imply that a large fraction of atoms leaves and/or accumulates within the filament to produce large relative concentration change. This, in turn, necessitates inclusion of strain energy in the filament formation modelling. In addition, the results reproduce non-linear *I-V* without the necessity of assuming the Poole-Frenkel type of electrical conduction or presence of a barrier at the oxide/metal interface.

INTRODUCTION

Resistive switching devices hold great promise for the next generation non-volatile solid state computer memories. The fabrication of these devices has been optimized in terms of functional material type and composition, electrode materials, deposition techniques, device design, and many other parameters to produce highly reliable devices. The most promising resistive switching devices appear to be Valence Change Memories (VCM) based on $TaO_x$.[1] In recent years, they entered the preproduction stage in major foundries.[2,3] However, the mechanism of the device operation and, in particular, the parameters of the small diameter filament responsible for the device conductance and switching are still a matter of dispute.

The conductance of VCMs is determined by the electroformation process which creates a small diameter conducting filament in initially uniform resistive functional oxide layer.[4] The filament is made of point defects comprising either oxygen vacancies and/or metal interstitials with both acting as donors in many functional oxides. Switching occurs by the ion motion within the filament opening a gap with low donor density in the filament. Given the variety of device designs (type of functional oxide and its oxygen content, electrode metals, thicknesses, and device sizes) and formation procedures, it is surprising that the *I-V* characteristics of formed devices are quite similar implying similar filament characteristics. Typically, the switching voltages range from 1-2 V while the switching currents are between 0.1 and 1 mA.[4–11] The simulations of the device *I-V* characteristics, however, arrived at the values of filament diameter and defect densities that vary widely. For example, the diameter of the filament was assumed to be as little as 5 nm[12] or as large as 600 nm[13] (this is four orders of magnitude difference of the filament cross sectional area). The width of the gap was between 1 nm[14] and 10 nm[15,16] while the donor concentration was typically in single atomic percent range of oxygen deficiency[17,18] with few studies considering the range of 10-20 %[14]. Such wide range of parameters was caused, in part, by *ad hoc* assumptions about many unknown material parameters. Another important consequence of such assumptions is the value of temperature simulated within the filament which varied from 500 K[14] to 3000 K[18]. It is apparent that the temperatures at high end of the range allow for very different phenomena to take place during switching compared to the low end of estimates.

This report attempts to narrow the range of the acceptable filament parameters (diameter, composition, and width of the gap in both the Low Resistance and High Resistance States (LRS and HRS, respectively) using fewest possible *ad hoc* assumptions. We limit the materials discussed to $TaO_x$ with electrodes forming ohmic contacts. The range of filament parameters was arrived at by using the values of current as a function of voltage and by the highest and lowest acceptable temperatures in HRS and LRS. The implications of the estimated filament composition are discussed in terms of energies of filament formation.

**Setting up the model and its evaluation criteria.**

The filament in LRS is frequently approximated in finite element models of resistive switching devices by a cylinder with a diameter $d$ and uniform composition $x$.[8–11,19–22] The composition here and the reminder of this report is defined as $x$ in TaO$_x$. This is equivalent to frequently used concentration of point defects ($c_D$ equal to the sum of oxygen vacancies and metal interstitials) as:

$$c_D(TaO_x) = (2.5 - x) \times 2.2 \times 10^{22} \ cm^{-3} \qquad (1)$$

HRS is created by moving donor ions in the direction of the applied field forming a gap with the composition $y$ and width $w$ close to anode. The reminder of the filament constitutes a "trunk" which we assumed has the same donor density as one in LRS. This is due to the length of the trunk being typically much larger than the width of the gap.[19,22,23] The diameter of the filament was assumed to be the same in LRS and HRS. The electrical conductivity of the functional oxide was assumed to be independent of the applied field with values as a function of temperature and composition discussed in Supporting Information. The detailed formulae for conductivity were obtained by fitting the experimental data of Bao et al.[24]

An important parameter of the model is the contact resistivity as in can control the temperature at the switching interface in the LRS. Our estimation listed below is an order of magnitude higher than that of Ascoli et al.[25] to account for the difference in contact material used: TiN versus Ta. The thermal boundary resistivity at TiN/TaO$_x$ interface was estimated at 5×10$^{-9}$ K m$^2$ W$^{-1}$ based on the data of Lyeo et al.[26] We have not included potential barriers at the electrode/oxide interfaces as there is no direct measurement of the barrier height as a function of composition or annealing and the shape of *I-V* can be reproduced without them. The diameter of the filament $d$, its composition $x$, as well as width of the gap $w$ and its composition $y$ were treated as adjustable parameters of our model with the veracity of the model assessed by the fit to selected experimental data summarized below.

The primary result of the simulations were the functional dependencies of current versus voltage for LRS and HRS. Experimental data show the characteristics in LRS as almost linear while the current in the HRS depends super-linearly on voltage.[12,16,23,27–30] The two experimental studies we use for direct comparison with simulation results here are reported by Ma *et al.* and Heisig *et al.*[23,27] These were selected as the authors have attempted to experimentally determine the parameters of the filament.

The other important characteristics of switching devices is the temperature distribution as it determines the rate of ion motion and switching speed. There is no direct measurement of the temperature within the filament. The most reliable estimates are based on injection of electrons from the filament into a p-type semiconductor by Yalon *et al.*[31] (T>1300 K), crystallization temperature of $HfO_x$ (Kwon *et al.*[32], T>800K), and *I-V* analysis (Sharma *et al.*[33], T=750K). Based on these results, we postulate that the lowest temperature allowing for switching is 800 K.

The activation energies for diffusion used in most simulations range from $1 < E_{diff} < 1.8$ eV [13,16,34,35] and are in general agreement with the direct measurements of diffusion rates.[36] Some simulations assumed lower activation energies and/or field-induced acceleration of diffusion.[12,14] Such low values were obtained as a fitting parameter and authors have not provided any independent justification. The consequences of $E_{diff} > 1$ eV are that to produce appreciable ion diffusion rates and switching dynamics, the temperatures within the filament must exceed 800 K. At the same time, the highest temperatures must stay below the melting temperatures of the electrodes and decomposition of the functional oxide. As the Ta-O phase diagram shows the liquid phase appearing at 1850 K, the estimated upper temperature limit was set to be at 1600 K.[37]

The maximum temperature within the filament depends on the dissipated power. It is worth noting that although the switching voltages can be quite different in different device structures, the power dissipated at the point of switching ranges from about 60 $\mu W$ to 180 $\mu W$ with the average of 110 $\mu W$.[5,23,27,38–42] The observation implies that the filaments should have similar structure in different devices switching occurs when the temperature reaches a particular value. The temperature distribution and maximum values discussed below, were obtained at the same dissipated power of 110 $\mu W$.

SIMULATION PROCEDURE

The TiN/TaO$_x$/TiN resistive switching devices were modeled using commercial COMSOL Multiphysics software package. The schematic diagram of the active volume of the device in the LRS is shown in the Figure 1(a) (the entire device structure is shown in Supporting Information). The as fabricated device has a 50 nm thick TaO$_{2.03}$ functional oxide layer and 40 nm thick TiN top and bottom electrodes. The electroformation process creates a cylindrical filament (marked by orange color in the figure) with composition TaO$_x$ and diameter $d$ both of which are adjustable parametrs in the simulation. The values of contact resistivities are marked with green lines for the contact between reduced oxide in the filament TaO$_x$ and TiN (6×10$^{-13}$ ohm m$^2$) and red lines for contact TiN/TaO$_{2.03}$ and TiN/TaO$_y$ (2×10$^{-12}$ ohm m$^2$). Figure 1(b) shows the device in the HRS with the gap in the filament next to the top electrode. The diameter and the composition of the filament in HRS is assumed to be the same as in LRS with the gap width $w$ and gap composition $y$ serving as parameters in the model. The circuit used for all simulation included a 12 kΩ load resistor. The *I-V*s were simulated using source voltage sweep from 0 V to 14 V in 1 ms. All *I-V* plots show current versus device voltage rather than source voltage. All temperature distributions were compared for a constant dissipated power of 110 $\mu W$.

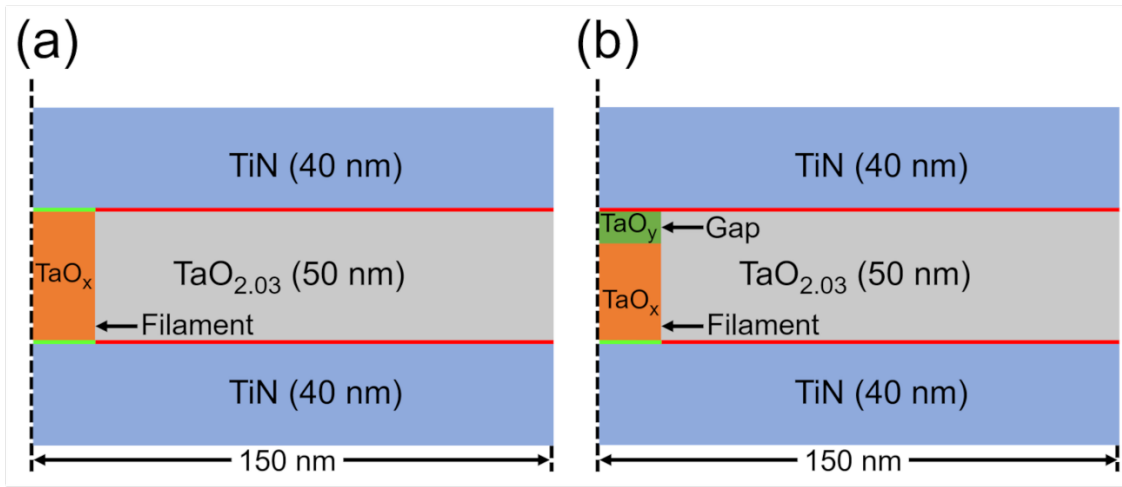

**Figure 1.** Axisymmetric cross-sectional view of the device, the black dash line representing the rotation axis. The green lines representing a lower contact resistance while the red lines are the position of higher contact resistance. (a) LRS state. (b) HRS state.

The model solved two coupled differential equations. The heat transport equation was:

$$\rho C_p \frac{\partial T}{\partial t} - \nabla \cdot (k_{th} \nabla T) = \vec{J} \cdot \vec{E} \qquad (2)$$

where $\rho$ denotes the mass density, $C_p$ the thermal capacity, $k_{th}$ the thermal conductivity, $T$ the temperature, $t$ the time, $\vec{J}$ the current density, and $\vec{E}$ the electric field. All material parameter values and boundary conditions are listed in the Supporting Information. The charge transport equation was:

$$\nabla(\sigma(x,T)\nabla\varphi) = 0 \tag{3}$$

where $\sigma(x,T)$ is the experimentally determined electrical conductivity and $\varphi$ is the electrostatic potential.

RESULTS AND DISCUSSION

**Low Resistance State**

The results of the *I-V* and temperature distribution simulations in the LRS along with two examples of experimental *I-Vs* are shown in Figure 2. Figure 2(a) shows the experimental *I-V* curve from Ma *et al.*[23] (dashed line) and Heisig *et al.* (dotted line).[27] Both show almost linear dependence characteristic of LRS. The resistance of Ma *et al.* device is larger than the one of Heisig *et al.* by an order of one magnitude. The Figure 2(b) shows the simulated *I-V* curves for the filament diameter $d = 16\ nm$ and the filament composition $x$ varying from 0.4 to 1.6. Besides the obvious increase of resistance, the curves show a change of functional dependence. For higher conductivity values with $x$ values 1.3 and below, the *I-V* curves are linear and the device resistance is between the values of Ma *et al.* and Heisig *et al*. In case of filament with higher oxygen content, the *I-V*s become superlinear due to the higher activation energy of electrical conductivity (Eq. S2). This limits the composition of the filament to below 1.3. Figure 2(c) shows the plot of the temperature distribution along the centerline of the filament. The distance scale starts at the interface with top electrode. For the filament composition $x = 0.4$, the maximum temperature is 998 K and occurs close to the interfaces with electrodes. This is due to filament conductance being higher than that of the contacts. With increasing oxygen content and decreasing conductivity of the filament, the maximum temperature location shifts to the middle of the oxide layer and its value reaches 1065 K at $x = 1.6$. Since the total dissipated power is constant, the rising oxygen content lowers the current, the dissipated power at contacts, and the temperature close to the interfaces. Composition change between 0.4 and 1.6 results in the temperature at the interfaces gradually dropping from 1065 K to 866K. Concomitantly, the temperature in the middle of the filament increased from 653 K to 1065 K even though the current dropped by over 30%. This decrease was compensated by the conductivity falling by almost 90%. Throughout all changes, the maximum temperatures stayed within acceptable 800 K – 1600 K range. However, the position of maximum temperature presents another limitation. Previous studies have shown that the gap in the filament was formed at the interface with the anode during the electroformation process.[23] This indicates that the temperature and ion diffusivity were the highest at this location or at least that both were more or less uniform throughout the filament. This is true for $x \leq 1.4$ but is not the case in filaments with the higher oxygen content. At $x = 1.6$ the difference of temperature in the middle of the filament and at the interface is about 200K indicating that ions are mobile and the gap should form in the center. The limits arrived at by analysis of Figures 2(b) and (c) agrees with the filament composition estimated by *Ma et al.*[23] ($TaO_{0.4}$) but not that of Heisig *et al.* ($TaO_{1.9}$-$TaO_{2.1}$).[27]

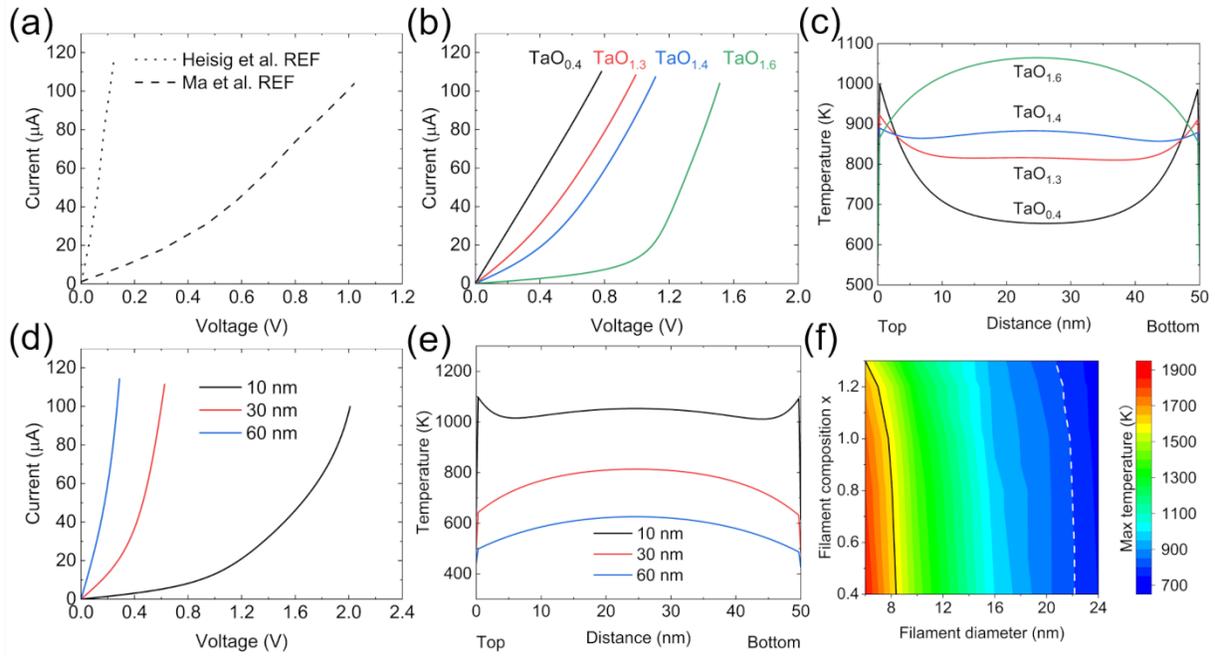

**Figure 2.** Characteristics of LRS. (a) Experimental *I-V* curves after Ma *et al.*[23] and Heisig *et al.*[27] (b) Simulated *I-V* curves and (c) the temperature line profile along the center of the filament as a function of filament composition for $d = 16\ nm$. (d) *I-V* curves for $x = 1.5$ as a function of filament diameter $d$. (e) Temperature along the line in the center of the filament ($x = 1.5$) as a function of filament diameter $d$. (f) Maximum temperature with the device in LRS as a function of filament composition $x$ and diameter $d$.

Figures 2(d) and (e) present the simulation results for the filament with fixed composition $x = 1.5$ and varying diameter. The *I-V* characteristics shown in Figure 2(d) indicate that quite obviously the conductance is increasing with the diameter and is accompanied by *I-V*s becoming more linear. This softens somewhat the previous result limiting the composition of the filament to below $x \leq 1.3$ allowing for somewhat higher oxygen content. However, increasing diameter rapidly lowers the maximum temperature within the filament (Figure 2(e)). The maximum temperature in device with 10 nm, 30 nm, and 60 nm filament diameter are 1098 K, 814 K and 626 K, respectively. This decrease is mostly due to lower dissipated power density and larger area of the filament surface. At the same time, the position of maximum temperature moves from interface to the middle. Since the maximum temperature is lower than the lowest acceptable value and its position is wrong, the filament composition $x = 1.5$ should be rejected even with larger filament diameters. The upper limit of filament composition should still be $x = 1.3$.

Figure 2(f) is the map of the maximum temperature within the LRS filament with different diameters and compositions. The black solid line on the left side of the figure is an isotherm corresponding to the upper limit of acceptable temperatures while the dashed white line represents the lower temperature limit. The dominant trend is the decrease of temperature with increasing diameter due to lower power density and more efficient lateral heat extraction. The dependence on composition is much less pronounced for $d \geq 12\ nm$ with the isotherms being almost vertical. The map indicates that the diameter of the filament should be between 6 and 22 nm.

The slight dependence of maximum temperature on composition can be understood in the following way. Lowering the composition, while keeping the diameter and dissipated power constant, results in slightly higher electrical conductance and higher current density. This increases the dissipated power density and temperature. The effect is mostly due to increasing relative importance of contact resistance. The dependence of maximum temperature on composition is stronger for small filament diameters with the difference of 264 K for $x = 1.3$ and $x = 0.4$ for $d = 6\ nm$.

**Modelling of High Resistance State**

Transmission electron microscopy (TEM) results from Ma *et al*. show that the filament diameter does not appreciably change during the set process.[23] Accordingly, we have assumed that the diameter remains the same in LRS and HRS with its size limits determined above for LRS need to be met in HRS as well. Figure 3 summarizes the results for the HRS as a function of the gap composition denoted here by $y$ and constant diameter $d = 16\ nm$, gap width $w = 5\ nm$, and filament composition $x = 1.3$. Panel 3(a) shows the simulated *I-V* curves as solid lines with the dashed and dotted lines corresponding to experimental HRS *I-V*s from reports by Ma *et al*.[23] and Heisig *et al*.[27] The plots are limited to a voltage range that did not allow for ion motion and permanent changes of the *I-V*. Both experimental curves exhibit strongly superlinear dependence of current on voltage with the *I-V* slope becoming almost vertical at higher voltages.

The simulations show increasing degree of nonlinearity with increasing oxygen content of the gap. For $y \geq 1.8$, the curves showed a region of negative differential resistance (NDR) which becomes more pronounced for higher $y$. NDR regions are expected in the *I-V* characteristics of devices using functional material with conductivity increasing with temperature[43] and are well documented and understood in structures based on $TaO_x$.[44,45] It is due to the thermally activated conductivity of $TaO_x$ and is more pronounced at higher oxygen content. With increasing composition, the overall resistance of the device is increasing moving the "knee" of the characteristics to higher voltages. When the $y$ increases to 1.9, the critical voltage increases to near 2.0 V while the voltage at maximum current is only 1.6 V. The NDR region has never been reported in the LRS and HRS of resistive switches. This is a very strong indication that the composition $x$ of the filament in LRS and the gap in the HRS is below 1.9. The gap composition of $y = 1.6$ results is device resistivities that are too low and the characteristics too linear. The acceptable gap composition $y$ then should be between 1.7 and 1.85. The effects of the filament diameter on the HRS are minimal as discussed in the Supporting Information.

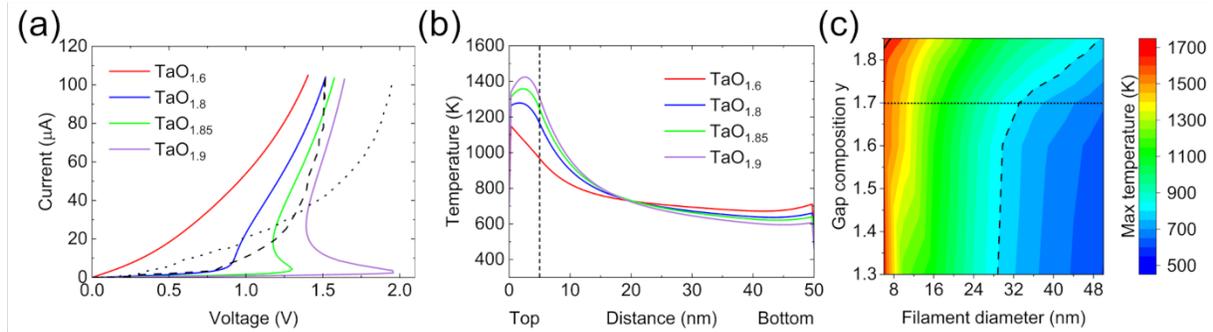

**Figure 3.** Dependence of HRS *I-V* characteristics on composition of the gap for width $w = 5\ nm$. (a) *I-V* curve (dashed line depicts experimental data of Ma *et al.*[23], dotted line is based on Heisig *et al.*).[27] (b) Temperatures along the centerline of device (gap extends from 0 nm to the dashed vertical line). (c) Map of the maximum temperature in the device.

Figure 3(b) shows the temperature distribution along the centerline of the filament as a function of the gap composition with other filament parameters the same as in Figure 3(a). The dissipated power is the same for all curves. The gap is located at the interface with the top electrode starting at 0 and extending to the dashed vertical line. The maximum temperature in the device always occurs within the gap and moves from the top interface to the middle of the gap with increasing $y$. It is increasing with oxygen content and remains in the acceptable temperature range.

The maximum temperature in HRS as a function of filament diameter and gap composition is shown in Figure 3(c) for constant width $w = 5\ nm$. The upper and lower limits of acceptable temperatures are marked by black solid and dashed lines (upper limit only shows in upper left corner of the figure). The dotted horizontal line at $y = 1.7$ is the lower limit of the gap composition arrived at in discussion of Figure 3(a). Both filament diameter $d$ and gap composition $y$ have significant influence on the maximum temperature. The diameter scale starts at 6 nm which is the lowest value allowed by the LRS. For the filament diameter range 6 to 8 nm and the constant diameter, the maximum temperature initially decreases and then increases with the gap composition. This is the same effect as encountered in the discussion of LRS. At low gap composition $y$, the maximum temperature is dominated by the Joule heating of the contact resistance. Accordingly, the maximum appears very close to the interface. At higher gap compositions, relatively more power is dissipated in the oxide with maximum temperature moving to the middle of the gap. The map does not impose any additional restrictions on the diameter of the filament beyond the limits obtained for the LRS.

In addition to the composition of the gap and filament diameter of the gap, the *I-V* characteristics could also be affected by gap width $w$ and the composition of the trunk. The composition of the trunk does not change the results much and can be neglected (estimates are discussed in the Supporting Information). The influence of gap width $w$ is summarized in Figure 4. The remaining parameters included composition of the trunk $x = 1.3$, gap composition $y = 1.8$, and filament diameter $d = 16$ nm.

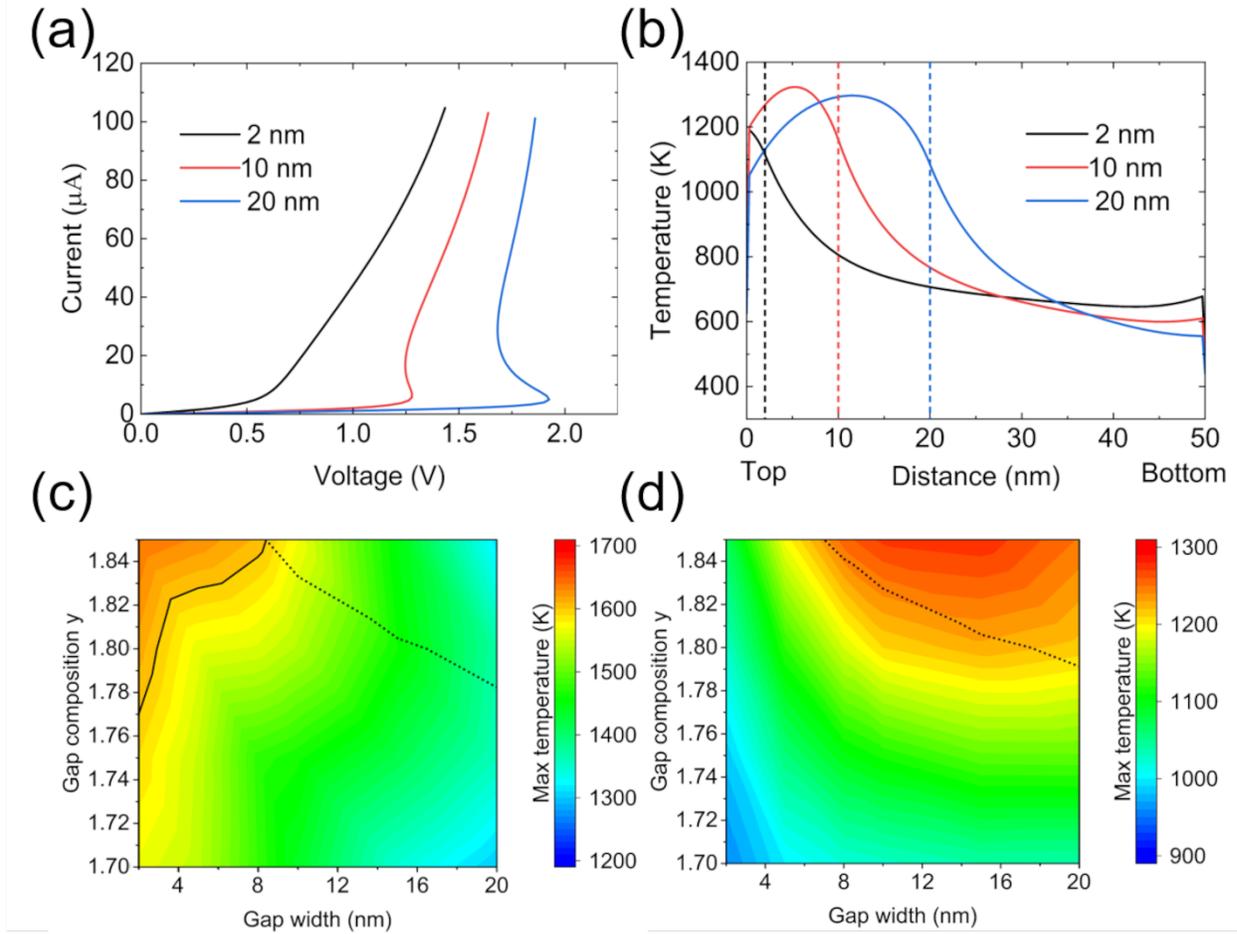

**Figure 4 (a) and (b).** Dependence of HRS on gap width and composition for filament diameter $d = 16\ nm$, trunk composition $x = 1.3$, and gap composition $y = 1.8$. (a) I-V curves. (b) Temperature line profile in the center of device, gap extends from 0 nm to the dashed line with corresponding color. **(c)** and **(d).** Maps of maximum temperature as a function of gap width and composition for filament diameter of (c) 6 nm and (d) 22 nm. Note the different temperature range in (c) & (d).

Figure 4(a) shows the *I-V* characteristics for different gap widths. The investigated range was from 2 to 20 nm. It is possible that the gap is narrower than 2 nm but the finite element models can be used only for sizes much larger than the interatomic distance. On the large side, 20 nm appeared as a reasonable limit as the diffusion times would be higher than that of interest for memory applications. For small values of $w$, the voltage as a function of current increases monotonically while for $w > 5$ nm it shows a local minimum corresponding to appearance of the NDR region. It should be noted that the *I-V* of the gap itself always exhibits NDR with the threshold voltage increasing with increase of its width. As the gap is connected in series with the fixed resistance of the trunk, the NDR becomes invisible for small values of the gap resistance.

Figure 4(b) shows the temperature distribution along the centerline of the filament with the edge of the gap shown as vertical dashed lines. The maximum temperature is reached close to the middle of gap except for small values of $w$ when the heat generated at the contact dominated the temperature distribution. The changes of the gap width have a relatively minor effect on the maximum temperature with the temperature changing from 1200 to 1300 K for width changing from 2 to 20 nm. The maps of the maximum temperature as a function of gap composition $y$ and filament diameter $d$ are shown in Figure 4(c) and Figure 4(d) for the filament diameter 6 and 22 nm. This covers the widest acceptable diameter range. The solid and dotted black lines correspond to 1600 K isotherm and the NDR region boundary, respectively, with NDR appearing for gap compositions above the dotted line.

For the small filament diameter (Figure 4(c)) the maximum temperature decreases with increasing gap width and decreasing gap composition (except for large $w$). The first effect is attributed to the decrease of current density, whereas the second is due to change of the temperature distribution. At lower $y$, the conductance of the gap increases and the power dissipated within it decreases. This lowers the temperature in the gap while increasing it elsewhere in the filament. For gap widths below 8.5 nm, the high gap composition are excluded due to exceeding the upper temperature limit. Also, for gap widths above 8.5 nm, devices with a high oxygen content in the gap are excluded due to the appearance of the NRD. For the gap composition $1.7 \leq y \leq 1.77$, there is no specific limitation on the gap width.

For the filament diameter at the upper limit of 22 nm (Figure 4(d)), the temperatures within the device are lower than for small diameter and do not limit the gap composition or width. Similar as for $d = 6$ nm, the maximum temperature falls with the decreasing composition but first rises and then falls with gap width consistent with the previously discussed results. The dotted line marking the appearance of NDR is very similar to one in panel (c).

The model presented above assumed abrupt changes of composition both at the boundaries of the filament. It is an oversimplification that can somewhat affect the values of the extracted parameters but the estimated effect of gradual changes is small.

DISCUSSION

The composition of the filament in resistive switching devices is typically discussed in terms of the concentration of oxygen vacancies with the typical concentrations assumed in the modelling reports in the $2 \times 10^{20} - 1.2 \times 10^{21}$ cm$^{-3}$ range.[12,16,17,46] This corresponds to $TaO_{2.49}$ - $TaO_{2.45}$ range in the notation used here. Only two modelling studies considered significantly lower oxygen content of $TaO_2$[14] and $TaO_{1.5}$[34]. The composition of the trunk of the filament estimated here is $x = 1.3$ or lower. In terms of relative oxygen concentration change the numbers above correspond to 0.3, 2, 20, 40, and 50 % of missing oxygen compared to fully oxidized TaO$_{2.5}$. Such large differences originate from very different parameters in commonly assumed dependence of electrical conductivity on composition and temperature:

$$\sigma(x,T) = \sigma_0(x)e^{-\frac{E_{act}(x)}{kT}} \quad (4)$$

In this report, we have used values of $\sigma_o$ and $E_{act}$ obtained by fitting experimental values of conductivity measured in wide composition and temperature range.[24] Our activation energies ranged from 0.34 eV for oxygen deficiency equal $3 \times 10^{21}$ cm$^{-3}$ to about 0 eV for oxygen deficiency $2.5 \times 10^{22}$ cm$^{-3}$. Most authors assumed activation energies corresponding to those of n-type silicon ($E_{act}$=0.05 eV for donor densities $n_D \sim 0$ and 0 for $n_D = 2 \times 10^{20}$ cm$^{-3}$) or lower.

In parallel, most authors, overestimated carrier mobility to compensate for using low carrier densities assumed to be the same as oxygen deficiency. The most direct test of the oxygen deficiency are spectroscopic methods. Their use in case of nanometer size filament is difficult but recently published data are in reasonable agreement with our assumed values. Analysis of High Angle Annular dark Field imaging estimated the composition of the filament at $x = 0.4$.[23] The X-ray Photo-Electron Emission Microscopy indicated $x = 2.0$,[27] and Scanning Nearfield Optical Microscopy gave $x = 1.2$.[47] The low temperature electrical conductivity measurements estimated $x \cong 1$.[48] While the above values appear to be scattered, they differ only by a factor of 4 and all are in the tens of percent range with the average of 50% or TaO$_{1.3}$.

Our composition estimate has very important consequences. The commonly assumed mechanism of electroformation process is the diffusion of oxygen out of the filament volume either laterally or vertically. To get $x \cong 1.3$ requires removal of large fraction of oxygen atoms from the volume of the filament and is bound to produce large tensile stresses and associated large elastic strain energy. The strain contribution to the total energy of formation can be neglected for oxygen deficiency of 1% but not for 50 % deficiency. In addition to elastic strain, one can expect plastic one as well resulting in formation of a depression on the surface of the top electrode. Since such depressions have not been reported, it is likely that the stress is compensated by in-diffusion of tantalum. This could be induced by either by stress itself or by an independent driving force such as Soret effect.[34,49]

One consequence of the $TaO_x$ composition in the filament estimated above is language we have used in most of this report. Specifically, we have used oxygen content, or the oxide composition rather than vacancy concentration as frequently done in the literature. The point defect concentrations or oxygen deficiency given above are mentioned only to compare our estimates with that in other reports. Generally, it is inappropriate to talk about oxygen vacancies in a material with composition much different from stoichiometric. The $TaO_{1.3}$ structure is not the same as that of $TaO_{2.5}$ with some oxygen sites left unoccupied.

An additional consequence of using electrical conductivity with large activation energies is the shape of *I-V*. Most experimental studies reported nonlinear *I-V* characteristics in HRS. Such nonlinearities have been modelled by assuming presence of Schottky barrier at the oxide-metal interfaces or by field dependent conductivity of $TaO_x$. While it is certainly possible that such effects exist, the thermal effects offer a simpler explanation. The linearity of *I-V* in LRS is due to low oxygen concentration throughout the filament and associated metallic conduction. Increase of temperature in the filament does not cause much change of conductivity. In HRS, the conductivity in the gap region exhibits significant activation energy and associated non-linear *I-V*.

The last comment, concerns the applicability of the results presented here to other material systems. While the limits of filament size and composition discussed here are not universal for resistive switches, it is likely that similar limits apply to another important material system i.e. metal/$HfO_2$/metal structures. This is due to similar dependence of $HfO_2$ conductivity on composition and temperature.[50]

CONCLUSIONS

A simple model of resistive switching devices consisting of a cylinder of reduced $TaO_x$ with or without a gap depleted of donor-type defects was used to reproduce *I-V* characteristics of High and Low Resistance States and the temperature distribution therein. The effect of four parameters was analyzed including: filament composition $x$, filament diameter $d$, gap composition $y$, and gap width $w$. Matching the shape of *I-V* with representative experimental data and constraining the maximum temperature to between 800 K and 1600 K, we have limited the filament diameter to between 6 and 22 nm. The composition of the trunk of the filament needs to be $x \leq 1.3$ to give linear LRS *I-V*. The same constraints restricted the composition of the gap to $1.7 \leq x \leq 1.85$. The filament composition agrees with the most recent experimental estimates. The model also demonstrates that the nonlinear *I-Vs* in HRS can be explained by Joule heating without necessity to invoke field-dependent conductivity or presence of Schottky barriers at the interfaces.

SUPPORTING INFORMATION

# Parametric study of "filament and gap" model of resistive switching in TaO$_x$-based devices.

Rongchen Li, Yang Bai, Marek Skowronski, Carnegie Mellon University, Pittsburgh PA 15213, USA

**Section 1. Details of COMSOL Simulations**

The electrical conductivity of the functional oxide TaO$_x$ was obtained by fitting experimental data of Bao *et al.*[1] using an expression:

$$\sigma(x, T) = \sigma_0(x) e^{-\frac{E_a(x)}{kT}} \tag{1}$$

The $\sigma_0(x)$ is the preexponential factor and the $E_a(x)$ is the activation energy, both of them being a function of the composition. The formulae for both are:

$$E_a(x) = \begin{cases} 0 \ (eV) & 0 \leq x < 0.667, \\ \exp\left(-\frac{18}{x+1} + 4.6\right) (eV) & 0.667 < x \leq 2.5. \end{cases} \tag{2}$$

$$\sigma_0(x) = \begin{cases} \frac{84000}{x+1} + 127491.6 \ (S/m) & 0 \leq x < 1.128, \\ \frac{1200000}{x+1} - 396944.4 \ (S/m) & 1.128 < x \leq 1.703, \\ \exp\left(\frac{48}{x+1} - 7\right) (S/m) & 1.703 < x \leq 2.5. \end{cases} \tag{3}$$

The experimental data covered the full range of compositions $0 \leq x \leq 2.5$ and wide range of temperatures (300-800 K). In simulations, the formulae above were used to extrapolate the conductivities to higher temperature. It should be pointed out that many of the previous simulations assumed conductivities with the pre-exponential factor much higher and activation energies much lower than ones adopted here.

Figure SI-1 shows the device design, materials used, and dimensions. Thermal and electrical boundary conditions are marked by green and red lines. Dotted green line denotes applied voltage while solid line denotes ground (V=0). Red dotted line at the periphery is the electrical insulation boundary. Both red and green dotted lines also correspond to thermal insulation boundaries. The dashed red line at bottom is the thermal boundary corresponding to set temperature of 300 K. The size of the slab was selected to represent much bigger wafer sizes used in the experiments.

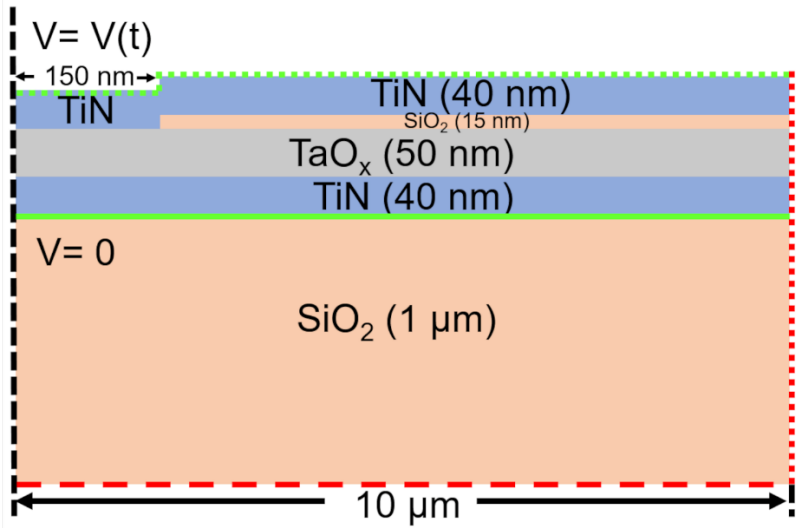

**Figure SI-1.** Boundary conditions used in the simulation.

The properties of bulk materials used in the simulation are listed in the Table S1.[2] Electrical conductivity of TaOx was a function of composition and temperature while all other properties were constant.

**Section 2. Additional simulations**

In the main text we have discussed the main effects of device parameters (filament composition $x$, filament diameter $d$, gap composition $y$ and gap width $w$) on the *I-V* characteristics and temperature distribution. Less important details which do not affect the conclusions reached in the main text are described here for completeness.

**Table S1. Material Properties**

| Materials | Electrical Conductivity S/m | Thermal Conductivity W m$^{-1}$ K$^{-1}$ | Heat Capacity J Kg$^{-1}$ K$^{-1}$ | Density Kg m$^{-3}$ | Relative Permittivity |
|---|---|---|---|---|---|
| TaOx | Eq.S1-3 | 0.6 | 174 | 8200 | 22 |
| TiN | 5·10$^6$ | 5 | 545 | 5220 | 4 |
| SiO$_2$ | 0 | 1.4 | 730 | 2200 | 3.9 |

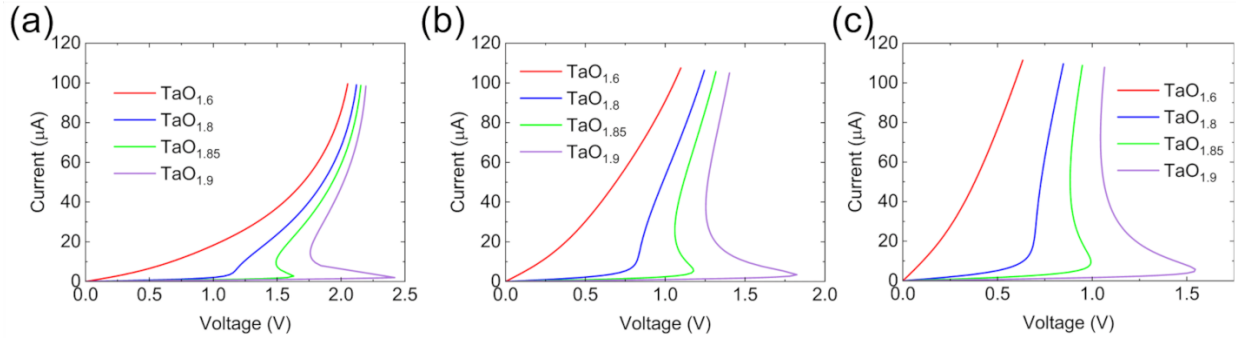

**Figure SI-2.** HRS *I-V* characteristics for different gap compositions and (a) 10 nm (b) 20 nm and (c) 30 nm diameter of the filament.

Figures 3 and 4 in the main use the appearance of Negative Differential Resistance (NDR) region in *I-V* as the criterion for limiting the composition and the width of the gap. In addition to these two parameters, the *I-V* can be affected by the diameter of the filament. Figure SI-2 below shows HRS *I-V* characteristics for different gap compositions and three values of the filament diameter. The composition of the filament trunk was fixed at $x = 1.3$. The three panels (a)-(c) show that the shape of *I-V* doesn't change much with the filament diameter $d$. All three *I-V* curves for $TaO_{1.6}$ and different filament diameters (red lines) are too linear compared to experimental *I-Vs* while $TaO_{1.9}$ curves (purple lines) exhibit NDR region.

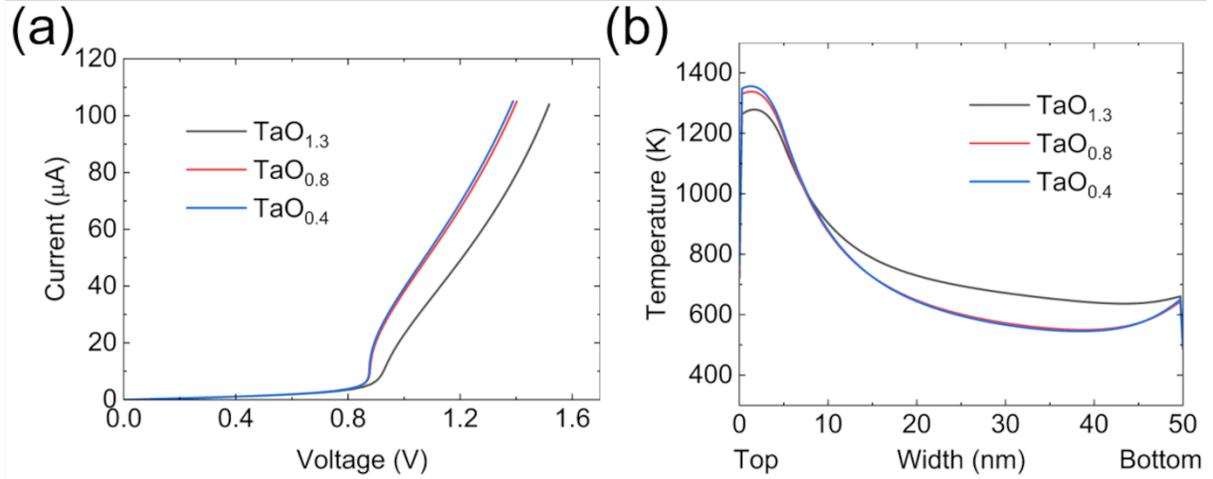

**Figure SI-3.** HRS *I-V* curve (a) and temperature distribution (b) for different filament trunk compositions.

Figure SI-3 shows the effect of changing the composition of the filament's trunk on the temperature distribution and *I-V* shape in HRS. The values of the remaining parameters (gap composition, filament diameter and gap width were fixed at 1.8, 16 nm, and 5 nm, respectively. The *I-V* curves in Fig. SI-3 have similar shape for all filament compositions. The change between $x = 1.3$ and 0.8 is minimal with the nonlinearity becoming somewhat less pronounced at low oxygen content of 0.4. Similarly small changes occur in the temperature distributions shown in Figure SI-3(b). The maximum temperatures for $x =$ 0.4, 0.8 and 1.3 are 1356 K, 1338 K and 1279 K respectively with the difference of less than 80 K. The temperature increase is mostly due to the increase of conductivity and current density at lower oxygen content. Both panels demonstrate that the characteristics of HRS model are dominated by gap in the filament and one can safely neglect the influence of filament trunk composition.